\documentclass[prd,twocolumn,nofootinbib,showpacs,superscriptaddress]{revtex4}

\usepackage{graphicx}
\usepackage{amsmath,amssymb,amsfonts,revsymb}
\usepackage{epsfig}

\def\to{\rightarrow}

\def\lsim{\mathrel{\rlap{\lower4pt\hbox{\hskip1pt$\sim$}}
    \raise1pt\hbox{$<$}}}
\def\gsim{\mathrel{\rlap{\lower4pt\hbox{\hskip1pt$\sim$}}
    \raise1pt\hbox{$>$}}}
\def\sqr#1#2{{\vcenter{\vbox{\hrule height.#2pt
         \hbox{\vrule width.#2pt height#1pt \kern#1pt
         \vrule width.#2pt}
         \hrule height.#2pt}}}}

\def\beq{\begin{equation}}
\def\eeq{\end{equation}}
\def\beqa{\begin{eqnarray}}
\def\eeqa{\end{eqnarray}}

\def\laq{\raise 0.4 ex \hbox{$<$}\kern -0.8 em\lower 0.62 ex\hbox{$\sim$}}
\def\gaq{\raise 0.4 ex \hbox{$>$}\kern -0.7 em\lower 0.62 ex\hbox{$\sim$}}

\begin{document}

\title{Observational constraints on modified gravity models and the Poincar\'e
                         dodecahedral topology}
\author{M.C. Bento}
\email{bento@sirius.ist.utl.pt} \altaffiliation{Also at Centro de
F\'{\i}sica Te\'orica de Part\'{\i}culas, Instituto Superior
T\'ecnico, Avenida Rovisco Pais, 1049-001 Lisboa}
\affiliation{%
Departamento de F\'{\i}sica, Instituto Superior T\'{e}cnico, Avenida
Rovisco Pais, 1049-001 Lisboa, Portugal}

\author{O. Bertolami}
\email{orfeu@cosmos.ist.utl.pt}
\affiliation{%
Departamento de F\'{\i}sica, Instituto Superior T\'{e}cnico,
Avenida Rovisco Pais, 1049-001 Lisboa, Portugal}

\author{M.J. Rebou\c{c}as}
\email{reboucas@cbpf.br}
\affiliation{%
Centro Brasileiro de Pesquisas F\'{\i}sicas  \\
Rua Dr.\ Xavier Sigaud 150 \\ 22290-180 Rio de Janeiro -- RJ, Brazil}

\author{N.M.C. Santos}
\email{n.santos@thphys.uni-heidelberg.de}
\affiliation{%
Institut f\"{u}r Theoretische Physik, Universit\"{a}t Heidelberg\\
Philosophenweg 16, 69120 Heidelberg, Germany}

\date{\today}

\begin{abstract}

We study observational constraints  on  models that account for the accelerated expansion
of the universe via infrared modifications to general relativity, namely the
Dvali-Gabadadze-Porrati  braneworld model as well as the Dvali-Turner  and Cardassian
models.  We find that significant constraints  can be placed on the parameters of each
model using type Ia supernovae data together with the baryon acoustic peak in the large
scale correlation function of the Sloan Digital Sky Survey of luminous red galaxies and
the Cosmic Microwave Background Radiation shift parameter data. Moreover, by considering
the Poincar\'e dodecahedral space as the circles-in-the-sky observable spatial topology,
we show that the detection of a such a non-trivial topology would provide relevant
additional constraints, particularly on the curvature parameter, for all models.
\end{abstract}



\maketitle

\section{Introduction}

Models where gravity is modified by soft very long-range corrections, normally inspired in
braneworld constructions, are an interesting approach to account for the recent
accelerated expansion of the universe, with no need for dark energy. One of the simplest
covariant modified-gravity models is based on the Dvali-Gabadadze-Porrati (DGP) braneworld
model \cite{DGPmodel}, as generalized to cosmology by Deffayet \cite{Deffayet2000}.  In
this model, gravity is altered at large distances by the slow leakage of gravity off our
 4-dimensional brane universe into the 5-dimensional bulk spacetime, leading to a
 modification of the Friedmann equation in a cosmological context. At
small scales, gravity becomes effectively bound to the brane and 4D gravity is recovered
to a good approximation. Crucially for our purposes, it was shown by Deffayet that the
model exhibits cosmological solutions with a self-accelerating phase
 at late times. An
interesting variation of this proposal has been suggested by Dvali and Turner
\cite{DvaliTurner2003} (hereafter referred to as DT model). Another possibility, also
originally motivated by extra dimensions physics, is the modification of the Friedmann
equation by the introduction of an additional nonlinear term proportional to $\rho^n$, the
so-called Cardassian model \cite{Freese:2002sq}\footnote{Other braneworld models could
also be considered, e.g. the model proposed in Ref. \cite{Sahni:2003}.}.

  We analyze current constraints on
the parameters of these models, as provided by the so-called \emph{gold} sample of 157
type Ia supernovae (SNe Ia)~\cite{Riess2004}, as well as the  baryon oscillation acoustic
peak (BAO) in the large scale correlation function of the Sloan Digital Sky Survey (SDSS)
\cite{Eisenstein:2005su} and the cosmic microwave background radiation (CMBR) shift
parameter \cite{Bond}. We consider a non-flat prior. Notice that joint SNe-CMBR-BAO
constraints have already been considered for the $\Lambda$CDM and quintessence models in
Ref. \cite{Ichikawa:2005} and for an f(R) modified gravity model in
\cite{Amarzguioui:2005}. While this work was in progress, the corresponding analysis for
the DGP model has appeared, see Ref. \cite{Maartens:2006yt}, where the authors conclude
that both flat DGP and $\Lambda$CDM models are within the 1 sigma contour, but the latter
provides a better fit to the data. In what concerns the DT and Cardassian models,
constraints from supernovae data alone have previously been studied, see
\cite{Bento:2004ym} and references therein.

Likewise for dark energy models, one expects the parameters of modified gravity models to be affected
by the geometry of the universe. The description of the universe as a metrical manifold, requires the
characterization of its geometry and its topology; hence, a key issue regarding our understanding of
the universe concerns its $3$--dimensional geometry and topology. Studies of the CMBR such as the ones
performed by the Wilkinson Microwave Anisotropy Probe (WMAP) allow for testing geometry, which is
related with the intrinsic curvature of the $3$--dimensional space. On the other hand, topology
concerns global properties of space such as its shape and size and, clearly, $3$--geometry restricts
but does not determine the topology of its spatial section. However, in a locally spatially homogeneous
and isotropic universe, the topology of its spatial sections determines the sign of its local
curvature~\cite{BernshteinShvartsman1980} and therefore dictates its geometry.

Different strategies and methodologies have been devised to probe a putative non-trivial topology of
the spatial sections of the universe (see, e.g. Refs.~\cite{CosmTopReviews,CCmethods} for reviews and
details on cosmic crystallographic methods). For instance, the so-called circles-in-the-sky method, is
based on the presence of multiple images of correlated circles in the CMBR maps~\cite{CSS1998}. In a
space with a detectable non-trivial topology, the last scattering sphere (LSS) intersects some of its
topological images along pairs of circles of equal radii, centered at different points on the LSS, with
the same distribution of temperature fluctuations, $\delta T$. These pairs of matching circles will be
imprinted on the CMBR anisotropy sky maps regardless of the background geometry or detectable topology
~\cite{CSS1998,CGMR05}. Hence, it follows that in order to probe observationally a non-trivial
topology, one should examine the full-sky CMBR maps in order to extract the correlated circles, and use
their angular radii and the relative position of their centers to probe a putative non-trivial topology
of the spatial sections of the observable universe.

In particular, it has been shown that the Poincar\'e dodecahedral space topology (see Section IV.A)
accounts for the low value of the CMBR quadrupole and octopole moments measured by first year WMAP
data~\cite{WMAP-Spergel:2003}, which has been confirmed by the most recent WMAP data
analysis~\cite{WMAP-Spergel:2006}, and fits the temperature two-point correlation
function~\cite{Luminet,Aurich1,Aurich2}. Recently, the Poincar\'e dodecahedral space~\cite{Luminet},
through the circles-in-the-sky method, has been considered as the observable spatial topology of the
universe in order to reanalyze the current SNe Ia plus X-ray gas mass fraction constraints on the
density parameters of matter ($\Omega_m$) and dark energy ($\Omega_{\Lambda}$) in the context of the
$\Lambda$CDM model~\cite{Previous}, with the result that it considerably reduces degeneracies. The
circles-in-the-sky method has also been used to place constraints on the parameters of the generalized
Chaplygin gas (GCG) model~\cite{Kamenshchik2001,Bilic2001,Bento2002}, as discussed in
Ref.~\cite{BBRS2006a}. In that work, by using both the Poincar\'e dodecahedral and binary octahedral
topologies, it has been shown that these spatial topologies through circles-in-the-sky could provide
additional constraints on the $A_s$ parameter of the GCG model as allowed by the SNe Ia observations.

Given these encouraging results it is natural to use this strategy to constrain the parameters of
modified gravity models as well. To this end, we will consider the Poincar\'e dodecahedral space
topology to reanalyze current constraints on the parameters of the DGP, DT and Cardassian models, in a
joint analysis with the observational constraints mentioned above, namely the \emph{gold} sample of SNe
Ia, as well as the baryon oscillation acoustic peak in the large scale correlation function of the SDSS
and the CMBR shift parameter.

\section{Modified-gravity models}

Modified gravity models explore the  possibility that there is no dark energy, and consider instead
that infrared modifications to general relativity exist on very large scales, accounting in this way
for the observed late time acceleration of the universe.

\subsection{Dvali-Gabadadze-Porrati model}

One of the simplest covariant modified-gravity models is based on
the DGP braneworld model~\cite{DGPmodel}, as generalized in
Ref.~\cite{Deffayet2000} to a FLRW brane in a Minkowski bulk.

In the DGP model, standard model gauge fields are confined to
a (3+1)D brane residing in an non-compact (4+1)D bulk, with different
scales of gravity on the brane and in the bulk. The gravitational
part of the action is  given by
\begin{align}
{S}\,=\, {M_{5}^3 \over 2} \,\int\,d^4x\,dw\,\sqrt{g^{(5)}}\, R_5\,
 +\,{M_{Pl}^2\over 2}\,\int\,
d^4x \sqrt{g}\,  R_4\, , \label{actionDGP}
\end{align}
where $M_5$ denotes de $5$D Planck mass, $M_{Pl}$ is the 4D Planck
mass, $g^{(5)}$ is the trace of the $5$D metric $g^{(5)}_{AB}$
$(A,B=0,1,2,...,4)$, $w$ is the extra spatial coordinate, $g$ the
trace of the $4$D metric induced in the brane,
$g_{\mu\nu}(x)~\equiv~g_{\mu\nu}^{(5)}(x, w=0)$, and where $R_5$, $R_4$ are
the $5$D and $4$D scalar curvatures, respectively. This
gravitational action coupled to matter on the brane leads to a
modified Friedmann equation, which can be written
as~\cite{Deffayet2000}
\begin{align}
H^2+\frac{k}{a^2}= \left(\sqrt{\frac{8\pi\,\rho}{3
M_{Pl}^2}+\frac{1}{4r_c^2}}+\frac{1}{2r_c}\right)^2 \,,
\label{dgpfried}
\end{align}
where
\begin{align}
r_c={M_{Pl}^2\over {2 M_5^3}}
\end{align}
is a length scale beyond which gravity starts to leak out into the
bulk.

Rewriting the above equation in dimensionless variables
$\Omega_x=\rho_x/\rho_{crit}$ with $\rho_{crit}=3
\,M_{Pl}^2\,H_0^2/8\pi$ and $\rho_x$ the energy density in the
component $x$  today, we get
\begin{align}
\left({{H} \over {H_0}}\right)^2=
\Omega_k(1+z)^2+\left(\sqrt{\Omega_m(1+z)^3+\Omega_{r_c}}+\sqrt{\Omega_{r_c}}\right)^2,
\label{leak}
\end{align}
where $H_0$ is the Hubble expansion parameter today  and $z$ is the
redshift, and we have taken into account that, at present, the
universe is matter dominated, hence $\rho \simeq \rho_{m}$.
Moreover, $\Omega_k=-{{k} \over {a_0^2 H_0^2}}$ is the present
curvature parameter and
\begin{align}
\sqrt{\Omega_{r_c}}= {1\over  2 r_c H_0}~.
\end{align}

The constraint equation between the various components of energy
density at $z=0$ is then given by
\begin{align}
\Omega_k+\left(\sqrt{\Omega_{r_c}+\Omega_m} + \sqrt{\Omega_{r_c}}\right)^2=1~.
\end{align}
It has been shown that the observed recent acceleration of the universe can be obtained
from the extra contribution to the Friedmann equation by setting the length scale $r_c$
close to the horizon size~\cite{Deffayet:2001,Deffayet:2002}.

\subsection{Dvali-Turner model}

Inspired in the above construction,  Dvali and Turner considered a more generic
modification of the Friedmann equation \cite{DvaliTurner2003}, hereafter referred
to as DT model
\begin{align}
\label{DTfried}
 H^2+\frac{k}{a^2}=\frac{8
\pi\,\rho}{3M_{Pl}^2}+\frac{1}{r_c^{2-\beta}}\left(H^2+\frac{k}{a^2}\right)^{\beta/2}~.
\end{align}
Notice that $\beta$ is the only parameter of the model: the case $\beta=1$ corresponds to the DGP
model, $\beta=0$ to the cosmological constant case, and  $\beta=2$ to a ``renormalization'' of the
Friedmann equation. A stringent bound follows from requiring that the new term does not interfere with
the formation of large-scale structure, $\beta \leq 1$, whereas the successful predictions of Big-Bang
nucleosynthesis impose a weaker limit on $\beta$, namely, $\beta \leq 1.95$. Moreover, it can be shown
that this correction behaves like dark energy in the recent past, with equation of state
$w_{eff}=-1+\beta/2$, and $w=-1$ in the distant future; moreover, it can mimic $w<-1$ without violating
the weak-energy condition \cite{DvaliTurner2003}.

The expression for the Hubble expansion as a function of redshift is then
\begin{align}
\left({{H} \over {H_0}}\right)^2
-2\sqrt{\Omega_{r_c}} \left[\left({{H} \over {H_0}}\right)^2  
- \,\Omega_k(1+z)^2\right]^{\beta/2}=  \nonumber \\
\Omega_m(1+z)^3+\Omega_k(1+z)^2 \;,
\label{Dtmodel}
\end{align}
where  $\Omega_{r_c}$ is now generalized to
\begin{align}
\sqrt{\Omega_{r_c}}=\frac{1}{2 (r_c \, H_0)^{2-\beta}} \;,
\end{align}
which means that the constraint between the various densities at $z=0$ is given by
\begin{align}
\Omega_m+\Omega_k+2\sqrt{\Omega_{r_c}}(1-\Omega_k)^{\beta/2}=1 \;.
\end{align}

\subsection{Cardassian model}

We will also consider the so called Cardassian
model~\cite{Freese:2002sq}, which explains the current acceleration
of the universe by a modification of the Friedmann equation
consisting basically in the introduction of  an additional term
proportional to $\rho^n$
\begin{align}
\label{FriedCmodel}
H^2= {{8 \pi}\over {3 M_{\rm Pl}^2}} \left( \rho + b \rho^n \right) - {{k} \over
{a^2}}~,
\end{align}
where $b$ and  $n$ are  constants, and we have added a curvature
term to the original Cardassian model. As in the previous cases, in
this model the universe is composed only of radiation and matter
(including baryon and cold dark matter) and the energy density
required to close the universe is much smaller than in standard
cosmology, so that matter can be sufficient to provide a flat (or
close to flat) geometry.

For $n < 1$ the second term becomes important if $z < {\cal O}(1)$; thereon it dominates
the Friedmann equation and yields $a \propto t^{2/3n }$ for ordinary matter, so
acceleration will occur if $n < 2/3$. There are two main motivations for the
introduction of the extra term, namely terms of that form typically when the universe is
embedded as a three-dimensional surface (3-brane) in higher dimensions \cite{Chung:1999zs}
or, alternatively, it may appear in a purely 4D theory due to an extra contribution to the
total energy density as would be the case if there were some unknown interactions between
matter particles \cite{Gondolo:2002fh}.

In a matter dominated universe, Eq.~(\ref{FriedCmodel}) can be
rewritten as
\begin{align}
\label{FriedCmodel2}
 \left({{H} \over {H_0}}\right)^2= \Omega_{m}
(1+z)^3+\Omega_k (1+z)^2 \nonumber \\
+ (1-\Omega_{m}-\Omega_k)(1+z)^{3n}~.
\end{align}
Notice that the case $n=0$ corresponds to the $\Lambda$CDM model.

Finally, it is worth pointing out that these models have been
thoroughly scrutinized from the observational point of view using
constraints from CMBR, SNe Ia and large scale structure. For recent
studies see, for instance,
Refs.~\cite{Sen:2003cy,Alcaniz-Pires:2004,Gong:2004sa,Zhu:2004ij,Fairbairn:2005ue,
Bento:2004ym,Maartens:2006yt,Elgaroy:2004ne,Guo-Zhu-Alcaniz-Zhang:2006,Alam:2006}.

\section{Observational constraints}

\subsection{Constraints from SNe Ia}

For our analysis, we consider the set of SNe Ia data recently compiled by Riess {\it et
al.}~\cite{Riess2004} known as the {\it gold} sample. This set contains 157 points: 143
points taken from the 230 Tonry {\it et al.}~\cite{Tonry:2003zg} data plus 23 points from
Barris {\it et al.}~\cite{Barris:2003dq} and 14 points discovered using
HST~\cite{Riess2004}. Various points where the classification of the supernovae was
uncertain or the photometry was incomplete have been discarded, thus increasing the
reliability of the sample. The data points in the {\it gold} sample are given in terms of
the distance modulus
\begin{align}
\mu_{\rm obs}(z) \equiv m(z) - M_{\rm obs}(z)~,
\end{align}
and the respective errors $\sigma_{\mu_{\rm obs}}(z)$, which already
take into account the effects of peculiar motions. The apparent
magnitude $m$ is related to the dimensionless luminosity distance
\begin{align}
D_L(z)&=\dfrac{1+z}{\sqrt{|\Omega_k|}}~ { S}\left(y(z)\right)~,
\end{align}
where  $ S(x)\equiv (\sin(x), \sinh(x), x)$ for $\Omega_k<0$, $\Omega_k >0$ and $\Omega_k=0$,
respectively, by
\begin{align}
m(z) = {\cal M} + 5 \log_{10} D_L(z) ~.
\end{align}
We have defined a new function
\begin{align}
y(z) \equiv \sqrt{|\Omega_k|}
\int_0^{z} \, \frac{H_0}{H(z')} \,\, dz'~.
\end{align}
The  $\chi^2$ is calculated from
\begin{align}
\chi^2_{SN} = \sum_{i=1}^n \left[ {{\mu_{\rm obs}(z_i) - {\cal M}' - 5 \log_{10}D_{L \rm
th}(z_i; \alpha_i)}\over{\sigma_{\mu_{\rm obs}}(z_i)}} \right]^2~,
\label{chisq2}
\end{align}
where ${\cal M}' = {\cal M} - M_{\rm obs}$ is a nuisance parameter, $\alpha_i$ are the
model parameters and $D_{L \rm th}(z;\alpha_i)$ is the theoretical prediction for the
dimensionless luminosity distance determined using the modified Friedmann equations.

\begin{figure}[t]
\centerline{\psfig{figure=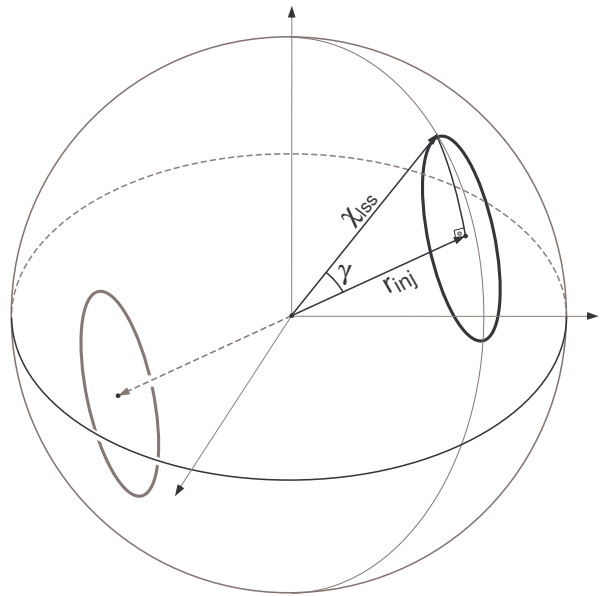,width=3.3truein,height=3.3truein,angle=0}} \caption{A
schematic illustration of two antipodal matching circles in the LSS. These pair of circles
come about in all globally homogeneous positively curved manifolds with a
circles-in-the-sky detectable topology. The relation between the angular radius $\gamma$,
angular sides $r_{inj}$ and $\chi^{}_{lss}$ is given by the following Napier's rule for
spherical triangles, $\cos \gamma = \tan r_{inj} \cot \chi^{}_{lss}$.}
\label{Ant_CinTheSky}
\end{figure}

\subsection{Constraints from the SDSS baryon acoustic oscillations}

In order to further remove degeneracies intrinsic to the distance fitting methods, it is interesting to
consider also the effect of the baryon acoustic peak of the large scale correlation function at $100
h^{-1}$~Mpc separation, detected by the SDSS team using a sample of LRG~\cite{Eisenstein:2005su}. The
position of the acoustic peak is related to the quantity
\begin{align}
{\cal A}&=\sqrt{\Omega_m}\left(\dfrac{H_0}{H(z_{lrg})}\right)^{1/3}
\,\left[\dfrac{1}{z_{lrg}\,\sqrt{|\Omega_k|}}~{ S}\left(y(z_{lrg})\right)\right]^{2/3}~,
\end{align}
which takes the value ${\cal A}_0=0.469 \pm
0.017$, and where $z_{lrg}=0.35$~\cite{Eisenstein:2005su}. We have neglected the weak
dependence of ${\cal A}_0$ on the spectral tilt. The baryon acoustic
peak is taken into account by adding the term
\begin{align}
\chi^2_{sdss}= \left(\dfrac{{\cal A}_{0}-{\cal A}}{\sigma_{{\cal A}}}\right)^2
\label{chisqpeak}
\end{align}
to the $\chi^2$, where $\sigma_{{\cal A}}$ is the error of ${\cal
A}_0$.

We should point out that there is a level of uncertainty in the measurement of ${\cal A}$
due to uncertainties essentially on $\Omega_m$ (notice that uncertainties on the baryon
density $\Omega_b$ are constrained by the CMBR and Big Bang nucleosynthesis to be smaller
than about $2 \%$). Also, one should notice that the baryon acoustic oscillations were
analyzed using a fiducial $\Lambda$CDM model and the full data set was compressed to a
constraint at a single redshift~\cite{Eisenstein:2005su}. As pointed out by Dick
\textit{et al}~\cite{Dick:2006ev}, the reduction of the data was intended to be valid for
the case of a $\Lambda$CDM model and robust for models with a constant equation of state,
but may give rise to significant systematic errors for the models we are considering.
Although a reanalysis of the baryon oscillation data in the context of these models would
no doubt be desirable, one may argue that substantial changes are not to be expected given
that the modifications of gravity we are considering are supposed to alter general
relativity only at the Gpc scale. The same can be said about topology given that it
affects only the low modes of the CMBR spectrum.

\begin{table*}[ht!]
\begin{tabular}{c c c c c c}
 \hline\hline
\hspace{1mm}Model\hspace{1mm} & \hspace{1mm}Parameters\hspace{1mm} &
\hspace{9mm}SN\hspace{9mm} &
 \hspace{7mm}SN+BAO\hspace{7mm} & \hspace{3mm}SN+BAO+CMBR\hspace{3mm}
 & SN+BAO+CMBR+T\\
\hline
 & $\Omega_m$&$0.46$ & $0.28$ &$0.28$& $0.29$\\
$\Lambda$CDM &$\Omega_k$& $-0.44$ & $0.033$ &$-0.003$&$-0.020$\\
 &$\chi^2$&$181.24$ & $183.76$&$183.93$&$184.44$ \\
 \hline
 & $\Omega_m$&$0.33$ & $0.27$&$0.27$& $0.28$\\
DGP &$\Omega_k$&$-0.56$ &$-0.32$ &$0.014$&$-0.021$\\
 &$\chi^2$&$181.36$ &$182.04$ &$190.53$&$192.34$ \\
 \hline
& $\beta$& $-10$   & $1.0$ &$0.26$&$0.23$\\
DT &$\Omega_m$& $0.49$   &  $0.27 $ &  $0.28$&$0.29$\\
 &$\Omega_k$& $0.032$   &  $-0.32$ &$-0.002$&$-0.02$ \\
 &$\chi^2$&  $180.55$   &  $182.04$ & $183.54$&$184.11$  \\
 \hline
& $n$& $-6.15$   & $0.33$  &$0.042$&$0.041$\\
Card &$\Omega_m$& $0.33$  & $0.27$  &$0.28$&$0.29$\\
 &$\Omega_k$& $0.33$  & $-0.76$  &$-0.003$&$-0.020$\\
 &$\chi^2$& $178.77$  & $182.08$  &$183.72$&$184.23$\\
\hline\hline
\end{tabular}
\caption{Best fit parameters for the $\Lambda$CDM, DGP, DT and Cardassian models for
different combinations of observational constraints (SN = SNe Ia gold sample, BAO = SDSS
baryon acoustic oscillations, CMBR = CMBR shift parameter and T~=Poincar\'e dodecahedral
space topology for $\gamma = 50^\circ \pm 6^\circ $).} \label{table:bestfits}
\end{table*}

\subsection{Constraints from the CMBR shift parameter}

It is expected that when the cosmological parameters are varied,
there is a shift in the whole CMBR angular spectrum, that is $\ell\to
{\cal R} \ell$, with the shift parameter $\cal R$ being given by~\cite{Bond}
\begin{align}
{\cal R}= \sqrt{\Omega_m\over \vert\Omega_k\vert}~ S(y(z_{lss})) ~,\label{shift}
\end{align}
where $z_{lss}=1089$~\cite{WMAP-Spergel:2003}. The results from CMBR (WMAP, CBI, ACBAR) data correspond
to ${\cal R}_0 = 1.716 \pm 0.062 $ (using results from Spergel et al. \cite{WMAP-Spergel:2003}). We
include the CMBR data in our analysis by adding
\begin{align}
\chi^2_{cmbr}=\left( {{\cal{R}}_0-{\cal R}\over \sigma_{\cal R}}\right)^2~,
\end{align}
to the total $\chi^2$ function, where $\cal R$ is computed for each model using Eq.
(\ref{shift}).
\begin{figure*}[hbt!]
\begin{center}
\includegraphics[height=5.9cm]{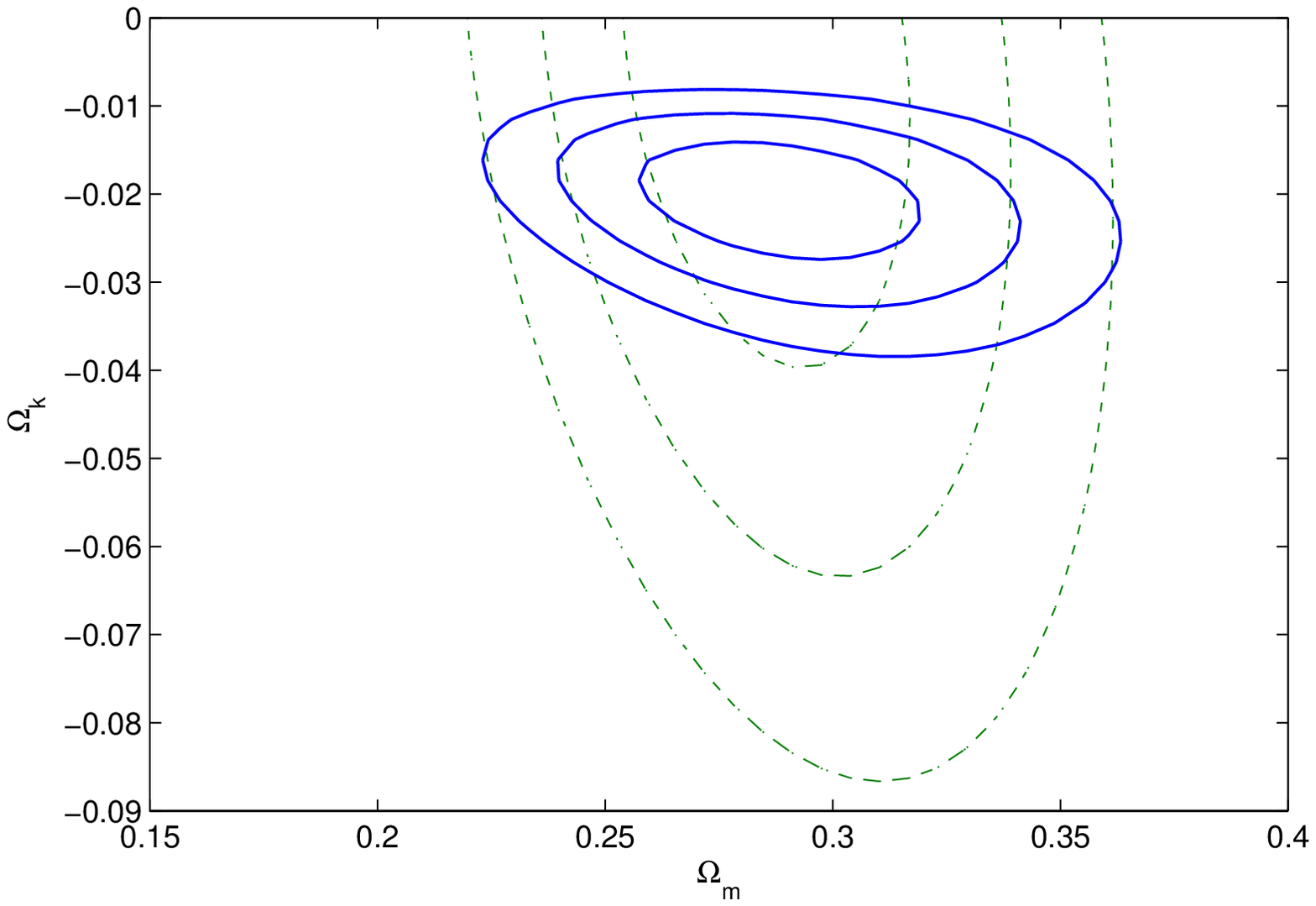}
\includegraphics[height=5.9cm]{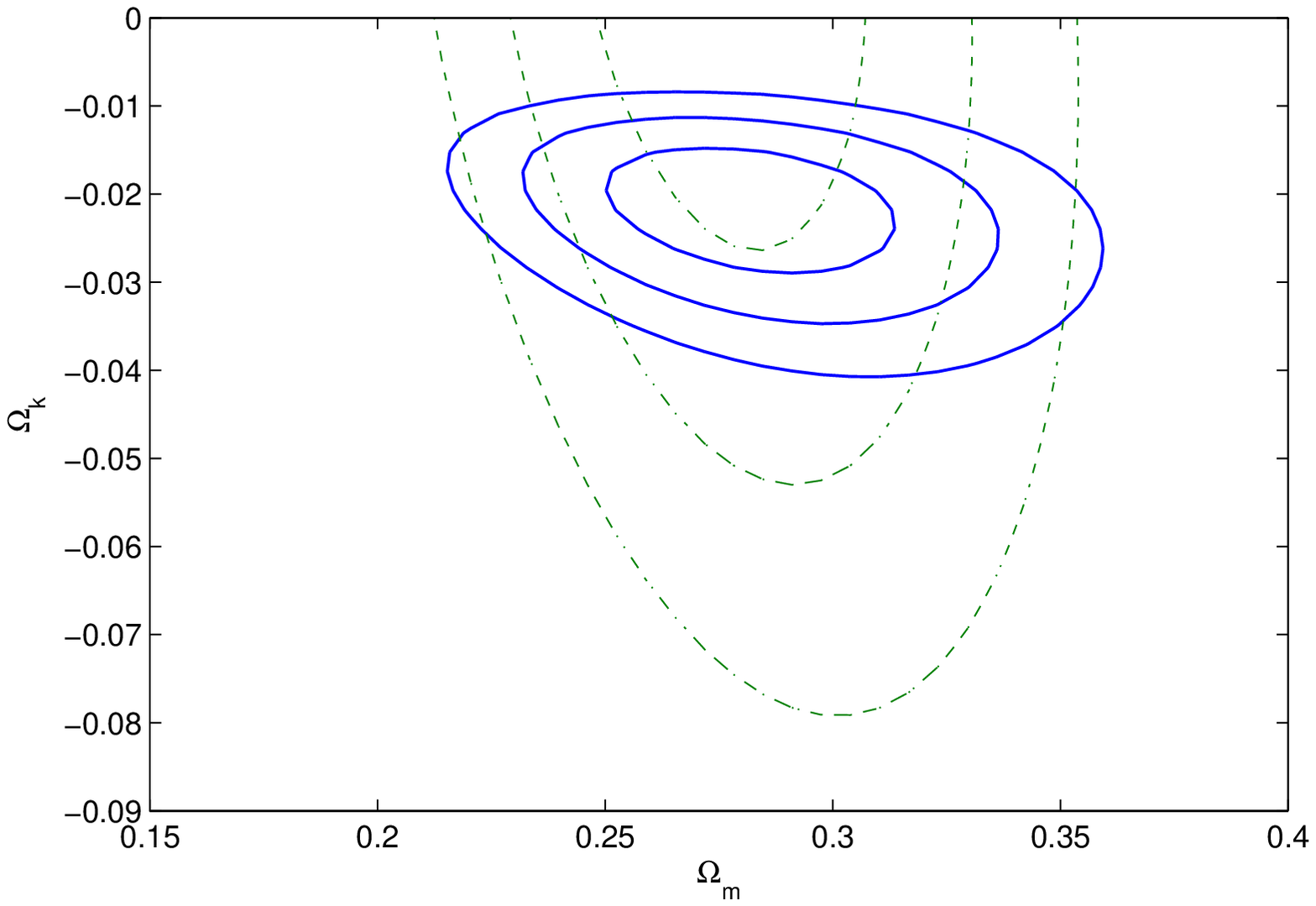}
\caption{\label{Fig:LambdaDGP}Confidence contours ($68.3\%$, $95.4\%$ and $99.7\%$) in the
$\Omega_m-\Omega_k$ plane for the $\Lambda$CDM (left) and DGP (right) models obtained with the SNe Ia
gold sample plus the SDSS acoustic peak data and CMBR shift parameter (dashed lines). Also shown are
the contours obtained assuming in addition a ${\cal D}$ space topology with $\gamma=50^\circ\pm
6^\circ$ (full).}
\end{center}
\end{figure*}
\section{Cosmic topology in brane cosmology}
%
In the framework of standard Friedmann--Lema\^{\i}tre--Robertson%
--Walker (FLRW) cosmology, the universe is described by a space-time manifold
$\mathcal{M}_4$ which is decomposed into $\mathcal{M}_4 = \mathbb{R} \times M_3$ and
endowed with a locally (spatially) homogeneous and isotropic  metric
\begin{equation}
\label{RWmetric} ds^2 = - dt^2 + a^2 (t) \left [ \frac{dr^2}{1-kr^2} + r^2 (d\theta^2 +
\sin^2 \theta \, d\phi^2) \right ] \,,
\end{equation}
where, depending on the sign of the constant spatial curvature $k$, the geometry of the
$3$--space $M_3$ is either Euclidean ($k=0$), spherical ($k=1$), or hyperbolic ($k=-1$).

Thus, since our $3$--dimensional space $M_3$ is chosen to be one of the following
simply-connected spaces, Euclidean $\mathbb{R}^3$, spherical $\mathbb{S}^3$, or hyperbolic
space $\mathbb{H}^3$, depending on the sign of the constant spatial curvature $k$, it is a
common misconception that the Gaussian curvature $k$ of $M_3$ is all one needs to
establish whether the $3$--space where we live in is finite or not. However, it is known
that the great majority of constant curvature $3$--spaces, $M_3$,  are multiply-connected
quotient manifolds of the form $\mathbb{R}^3/\Gamma$, $\mathbb{S}^3/\Gamma$, and
$\mathbb{H}^3/\Gamma$, where $\Gamma$ is a fixed-point free group of isometries of the
corresponding covering space. Thus, for example, for the Euclidean geometry besides
$\mathbb{R}^{3}$ there are 6 classes of topologically distinct compact and orientable
spaces $M_3$ that can be endowed with this geometry, while for both the spherical and
hyperbolic geometries there is an infinite number of non-homeomorphic (topologically
inequivalent) manifolds with non-trivial topology that admit these geometries. On the
other hand, since the ultimate spatial topology has not yet been determined by
cosmological observations, our $3$--dimensional space may be any of these possible
quotient manifolds.

Quotient manifolds are compact in three independent directions, or compact in two or at
least one independent direction. In compact manifolds, any two given points may be joined
by more than one geodesic. Since the radiation emitted by cosmic sources follows
geodesics, the immediate observational consequence of a nontrivial
detectable spatial non-trivial topology%
\footnote{The extent to which a non-trivial topology may be detected was discussed in
Refs.~\cite{TopDetec}.}
of $M_3$ is that there will be multiple images of either cosmic objects or specific spots
on the CMBR. At very large scales, the existence of these multiple images (or pattern
repetitions) is a physical effect that can be used to probe the $3$-space topology.

In  5D braneworld models, the universe is described by a $5$-dimensional metrical orbifold
(bulk) $\mathcal{O}_5$ that is mirror symmetric ($\mathbb{Z}_2$) across the 4D brane
(manifold) $\mathcal{M}_4$. Thus, the bulk can be decomposed as $\mathcal{O}_5 =
\mathcal{M}_4 \times E_1 =  \mathbb{R} \times M_3 \times \mathbb{E}_1$, where $E_1$ is a
$\mathbb{Z}_2$ symmetric Euclidean space, and  where $\mathcal{M}_4$ is endowed with a
Robertson--Walker metric Eq. (\ref{RWmetric}), which is recovered when $w=0$ for the extra
non-compact dimension. In this way, the multiplicity of possible inequivalent topologies
of our $3$--dimensional space, and the physical consequences of a non-trivial detectable
topology of $M_3$ (possible multiple images of discrete cosmic sources, circle-in-the-sky
on the LSS) is brought on the braneworld scenario.

\subsection{Poincar\'e Dodecahedral Space Topology}

The Poincar\'e dodecahedral space $\mathcal{D}$ is a $3$-manifold of the form
$\mathbb{S}^3/\Gamma$ in which $\Gamma=I^\star$ is the binary icosahedral group of order
$120$. It is represented by a regular spherical dodecahedron ($12$ pentagonal faces) along
with the identification of the opposite faces after a twist of $36^\circ$. Such a space is
positively curved (k=1, $\Omega_k<0)$, and tiles the $3$--sphere $\mathbb{S}^3$ into $120$
identical spherical dodecahedra.

The observed values of the power measured by WMAP of the CMBR quadrupole ($\ell=2$) and
octopole ($\ell=3$) moments, and the sign of the curvature density $\Omega_{k}=- 0.02
\pm\, 0.02$ reported by first year WMAP data analysis team~\cite{WMAP-Spergel:2003}, which
has been reinforced by the three-year WMAP observations (cf. Table~11 of
Ref.~\cite{WMAP-Spergel:2006}), have motivated the suggestion of the Poincar\'e
dodecahedral space topology as an explanation for this observed large-angle anomaly in the
CMB power spectrum~\cite{Luminet}. This observation has sparked the interest in the
dodecahedral space, which  has been examined on its various
features~\cite{Cornish,Roukema,Aurich1,Gundermann,Aurich2}. In particular, it turns out
that a universe with the Poincar\'e dodecahedral space section squares with WMAP data in
that it accounts for the suppression of power at large angle observed by
WMAP~\cite{WMAP-Spergel:2003,WMAP-Spergel:2006}, and fits the WMAP temperature two-point
correlation function~\cite{Aurich1,Aurich2}, retaining the standard FLRW background for
local physics. Notice however that a preliminary search of the antipodal matched circles
in the WMAP sky maps predicted by the Poincar\'e model has failed~\cite{Cornish}. A second
search of the correlated circles for the $\mathcal{D}$ space was not conclusive
either~\cite{Aurich3}. This absence of evidence of correlated circles may be due to
several causes, such as the Sunyaev-Zeldovich effect, lensing and the finite thickness of
the LSS, as well as possible systematics in the removal of the foregrounds, which can
damage the topological circle matching. Thus, it is conceivable that the correlated
circles may have been overlooked in the CMBR sky maps search~\cite{Aurich1}.

Regarding the  compatibility of the Poincar\'e dodecahedral space topology with string
theory, it has been shown that, on account of the Adams-Polchinski-Silverstein conjecture
on the instability of non-supersymmetric AdS orbifold,  the Poincar\'e dodecahedral space
topology cannot arise as a model for the spatial sections of accelerating braneworld
cosmological models in the framework of string theory~\cite{McInnes:2004}. However, while
the braneworld paradigm has often been referred to as "string-inspired," the models we are
considering are not committed to that premise.

\begin{figure*}[ht!]
\begin{center}
\includegraphics[height=5.9cm]{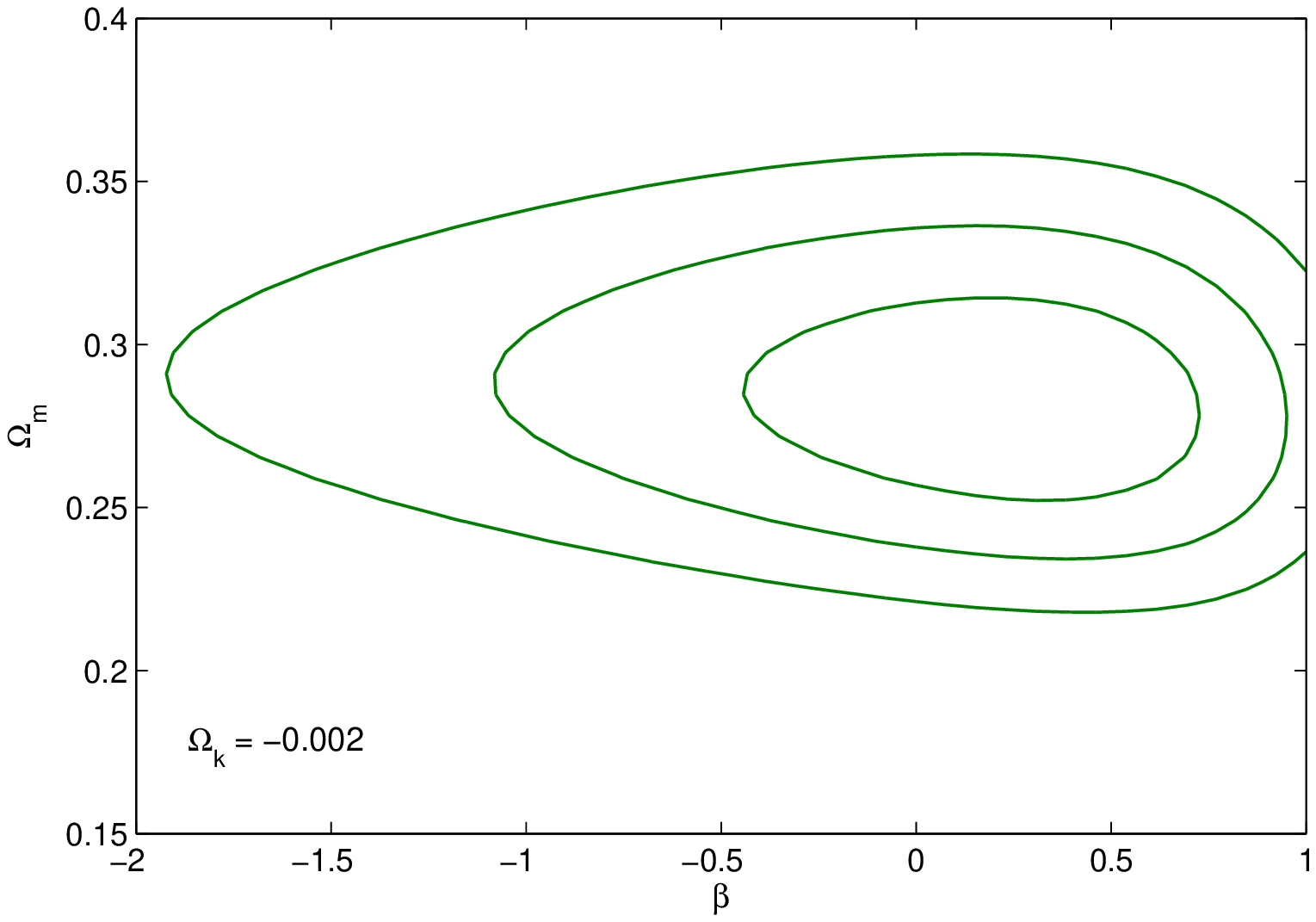}
\includegraphics[height=5.9cm]{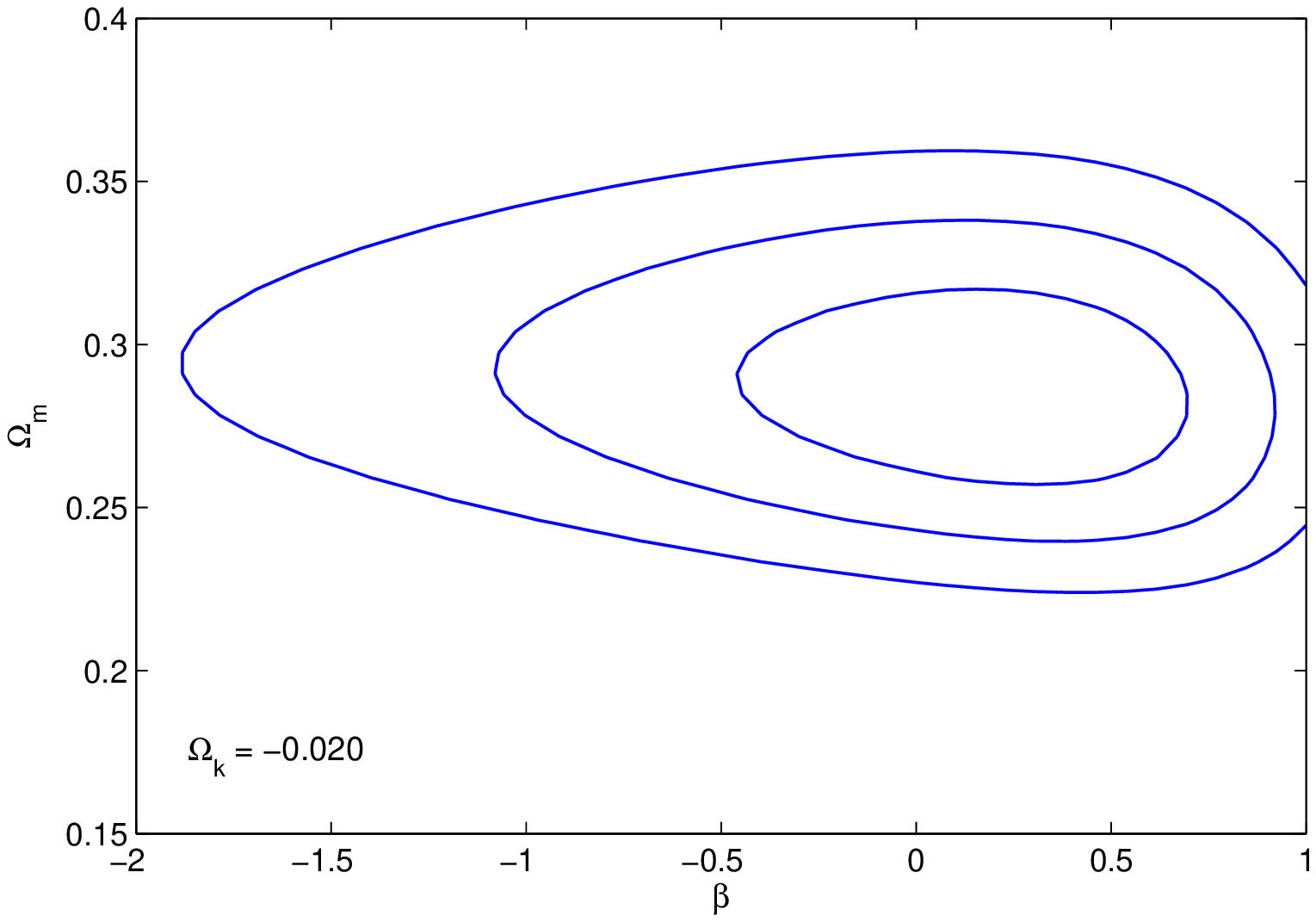}
\includegraphics[height=5.9cm]{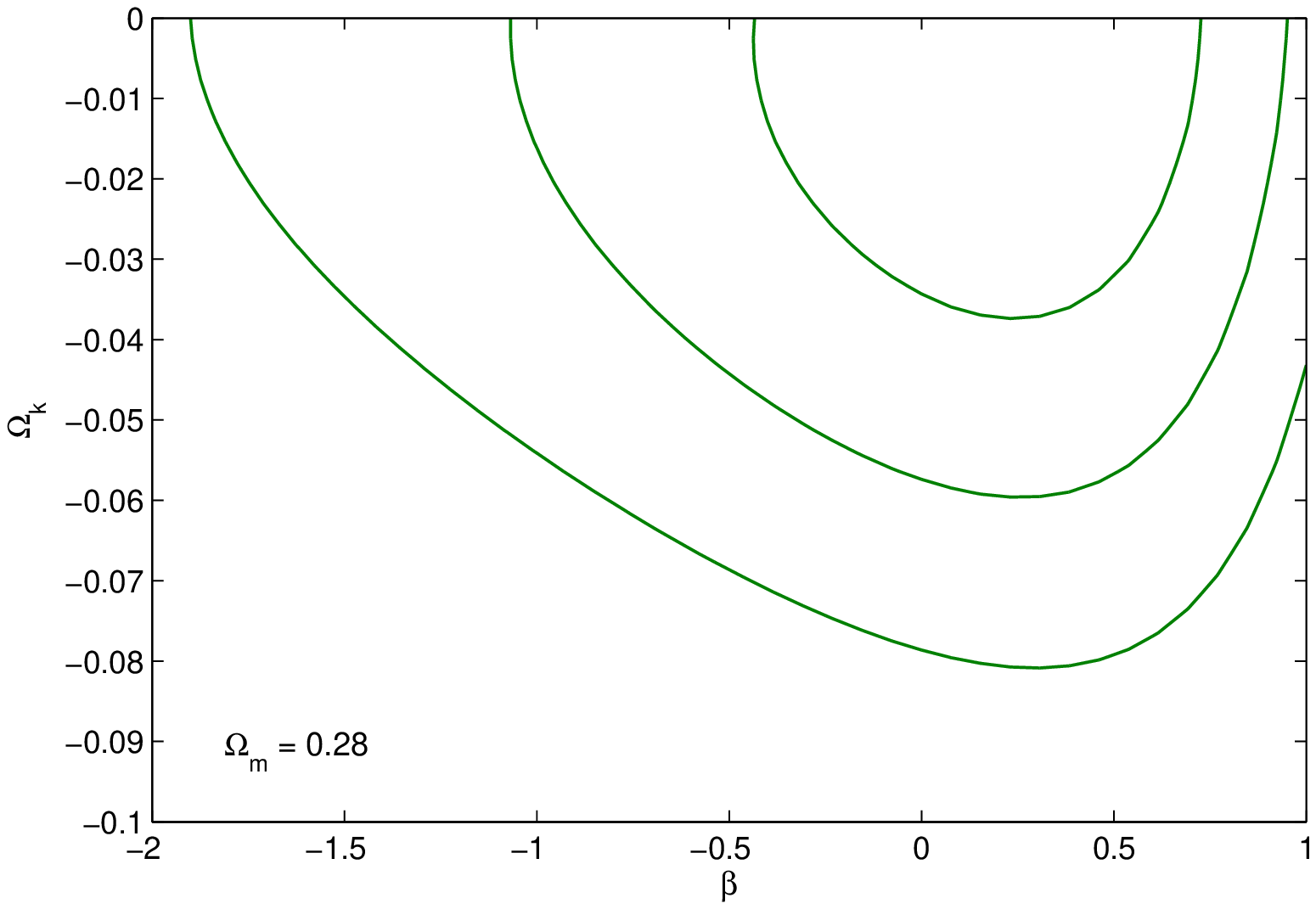}
\includegraphics[height=5.9cm]{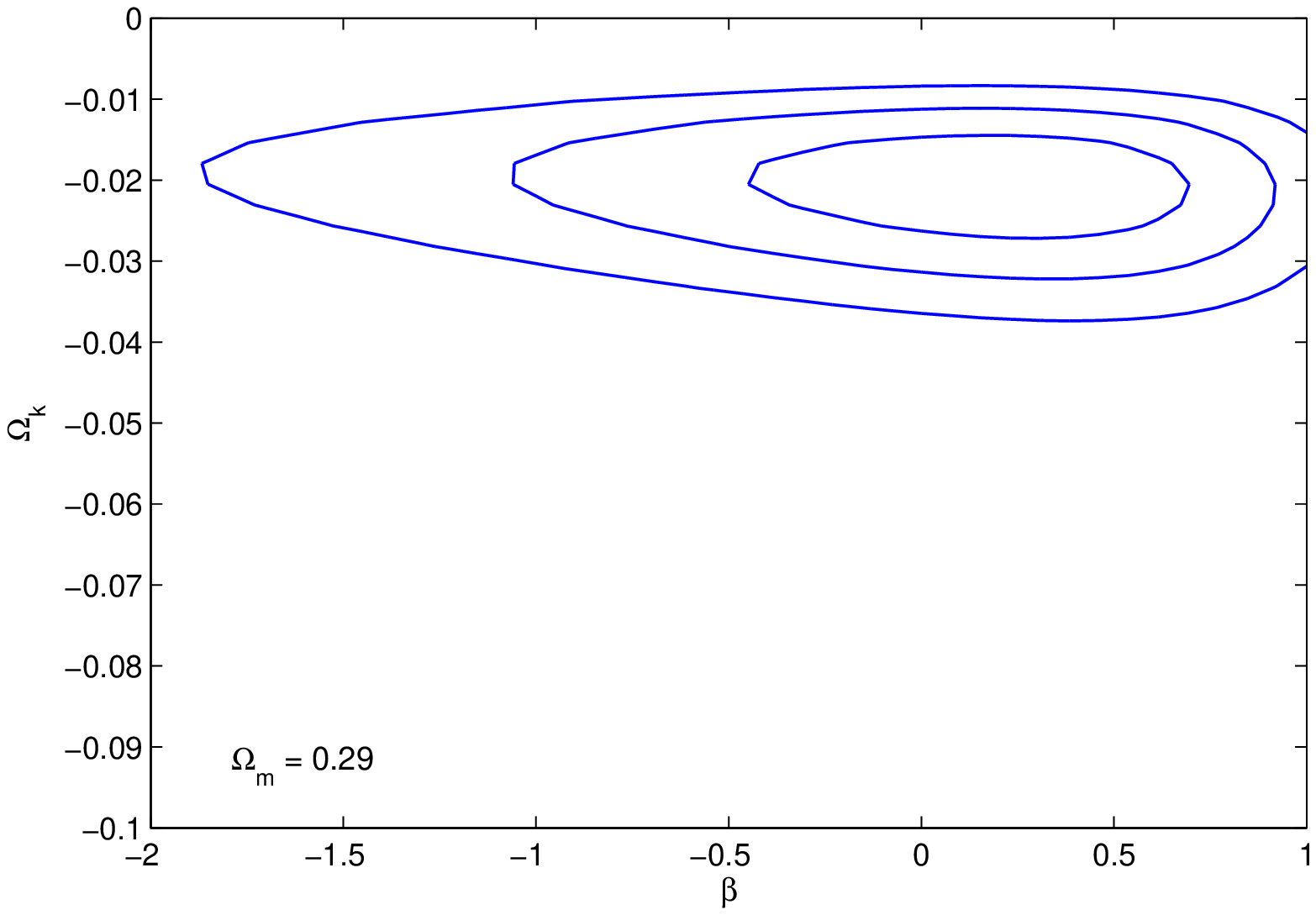}
\caption{\label{Fig:DT}Confidence contours ($68.3\%$, $95.4\%$ and $99.7\%$) for the DT
model, in the $\Omega_m-\beta$ (top) and $\Omega_k-\beta$  (bottom) planes,  obtained with
the SN + SDSS + CMB data (left) and including also a ${\cal D}$ space topology with
$\gamma=50^\circ \pm 6^\circ$ (right). $\Omega_k$ or $\Omega_m$ are fixed at the value
that minimizes the $\chi^2$.}
\end{center}
\end{figure*}
\subsection{The Circles-in-the-Sky Method}

An important class of constant curvature positively-curved $3$-spaces with a non-trivial
topology is comprised by the \emph{globally homogeneous} manifolds. These manifolds
satisfy the topological principle of (global) homogeneity, in the sense that all points in
$M_3$ are topologically equivalent. In particular, in these spaces the pairs of matching
circles will be antipodal, as shown in Figure~\ref{Ant_CinTheSky}.

The Poincar\'e dodecahedral space $\mathcal{D}$ is globally homogeneous, and give rise to
six pairs of antipodal matched circles on the LSS, centered in a symmetrical pattern as
the centers of the faces of the dodecahedron. Figure~\ref{Ant_CinTheSky} shows a pair of
these antipodal circles. Clearly the distance between the centers of each pair of these
correlated circles is twice the radius $r_{inj}$ of the sphere inscribed in $\mathcal{D}$.

It then follows from the use of trigonometric relations (known as Napier's rules) for the
right-angled spherical triangle shown in Fig.~\ref{Ant_CinTheSky} gives origin to a
relation between the angular radius $\gamma$, the angular sides $r_{inj}$ and radius
$\chi^{}_{lss}$ of the LSS, namely
\begin{equation}
\label{Chigamma} \chi^{}_{lss} = \tan^{-1} \left[\,\frac{\tan r_{inj}}{ \cos \gamma}\,
\right] \;,
\end{equation}
where $r_{inj}$ is a topological invariant, equal to $\pi/10$ for the the space
$\mathcal{D}$, and the distance $\chi^{}_{lss}$ to the
origin 
is given by
\begin{equation}
\label{ChiLSS} \chi^{}_{lss}=y(z_{lss})~.
\end{equation}
Eq.~(\ref{ChiLSS}) makes apparent that $\chi^{}_{lss}$ depends on the cosmological
scenario. Moreover, Eq.~(\ref{Chigamma}) with $\chi^{}_{lss}$ given by Eq.~(\ref{ChiLSS})
together with the ratio $H_0/H$ for each modified-gravity model yield a relation between
the angular radius $\gamma$ and the cosmological parameters of each model. Thus, they can
be used to set bounds (confidence regions) on model parameters. To quantify this, we
consider a typical angular radius $\gamma = 50^\circ$ estimated in Ref.~\cite{Aurich1} for
the Poincar\'e dodecahedral space and, since the measurements of the radius $\gamma$ do
involve observational uncertainties on the model parameters from the detection of the
spatial topology, we take into account these uncertainties; in order to obtain
conservative results, we consider $\delta \gamma \simeq 6^\circ$, which is the scale below
which the circles are blurred for the Poincar\'e dodecahedral space case~\cite{Aurich1}.

%

\begin{figure*}[ht!]
\begin{center}
\includegraphics[height=5.9cm]{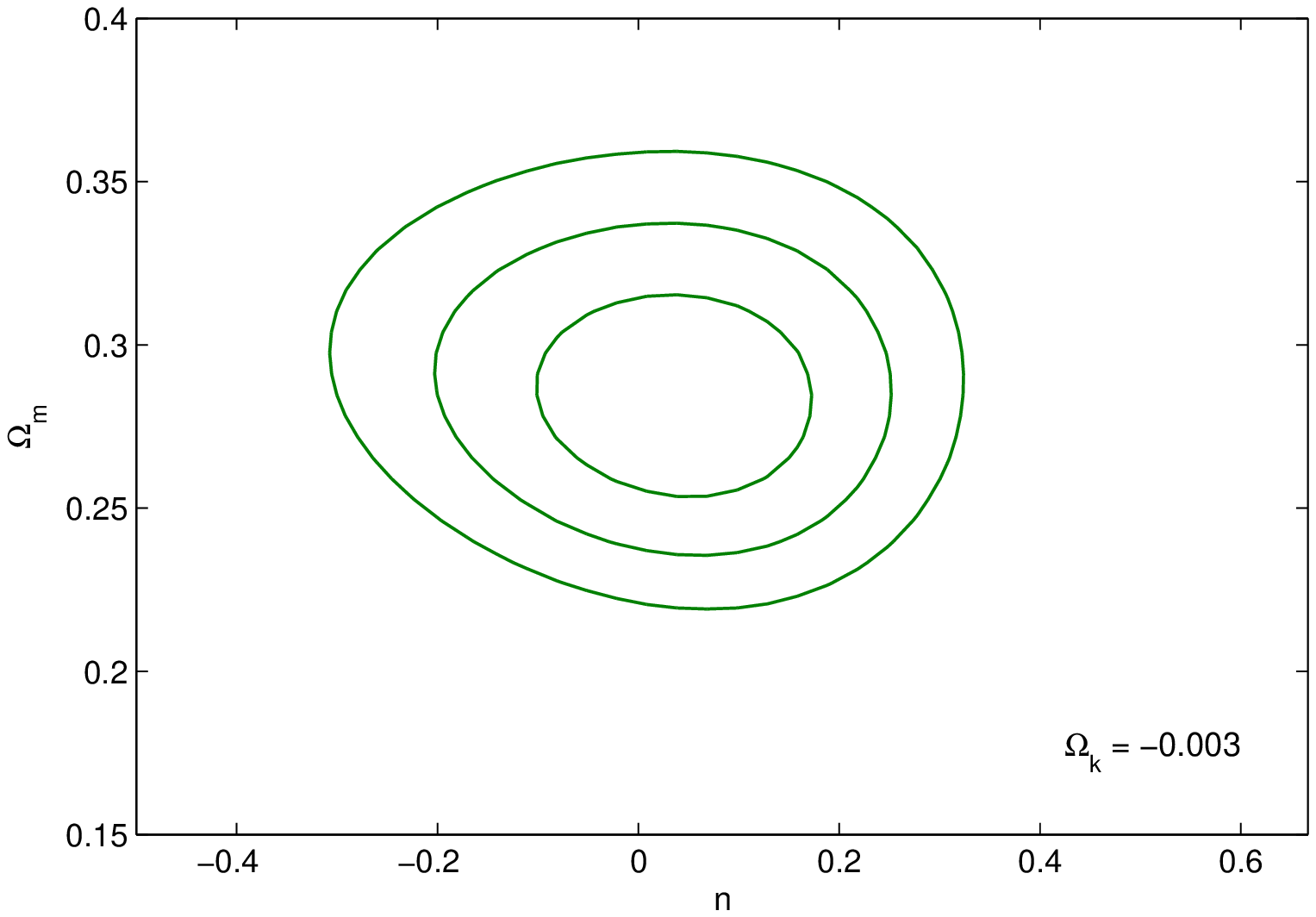}
\includegraphics[height=5.9cm]{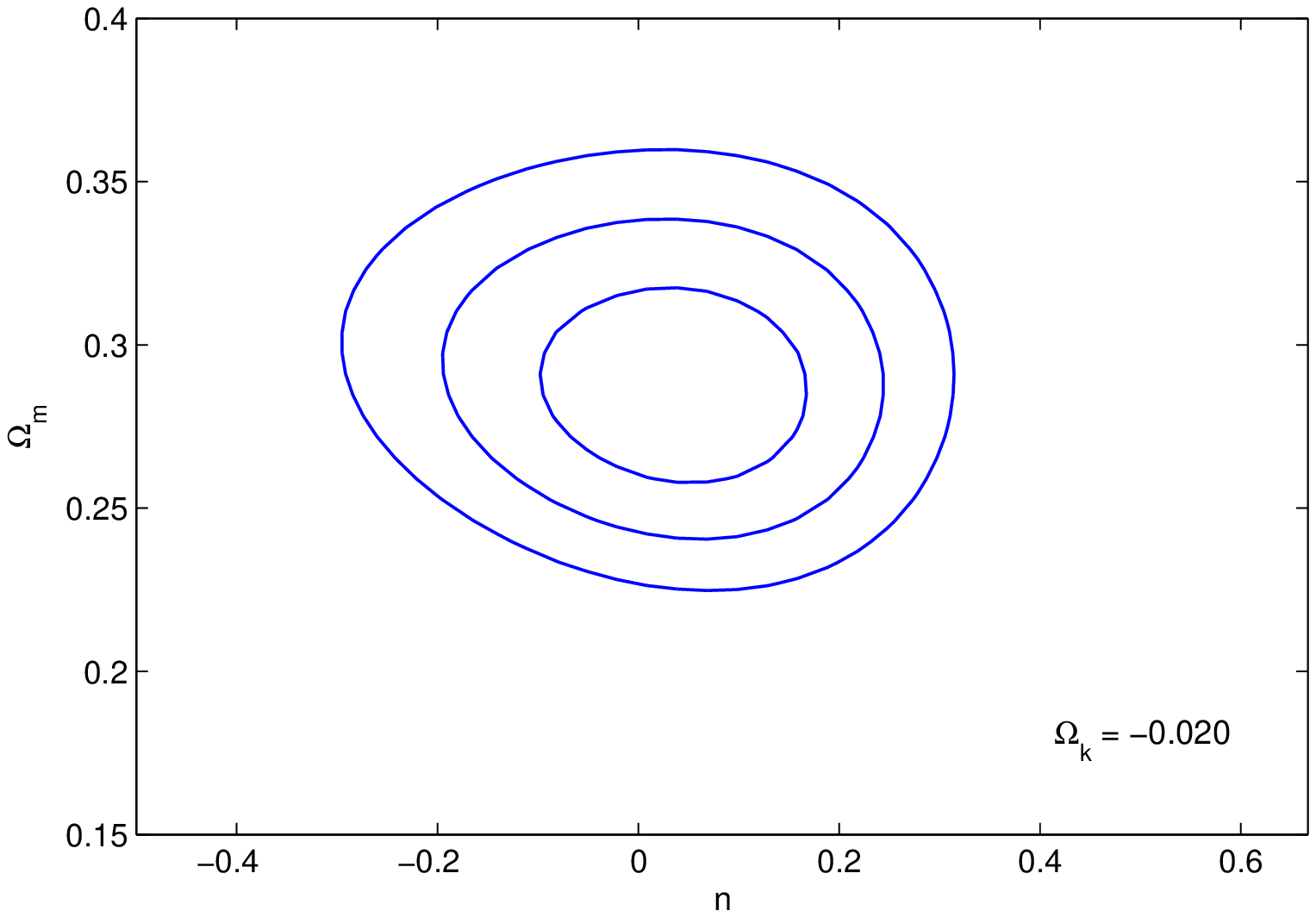}
\includegraphics[height=5.9cm]{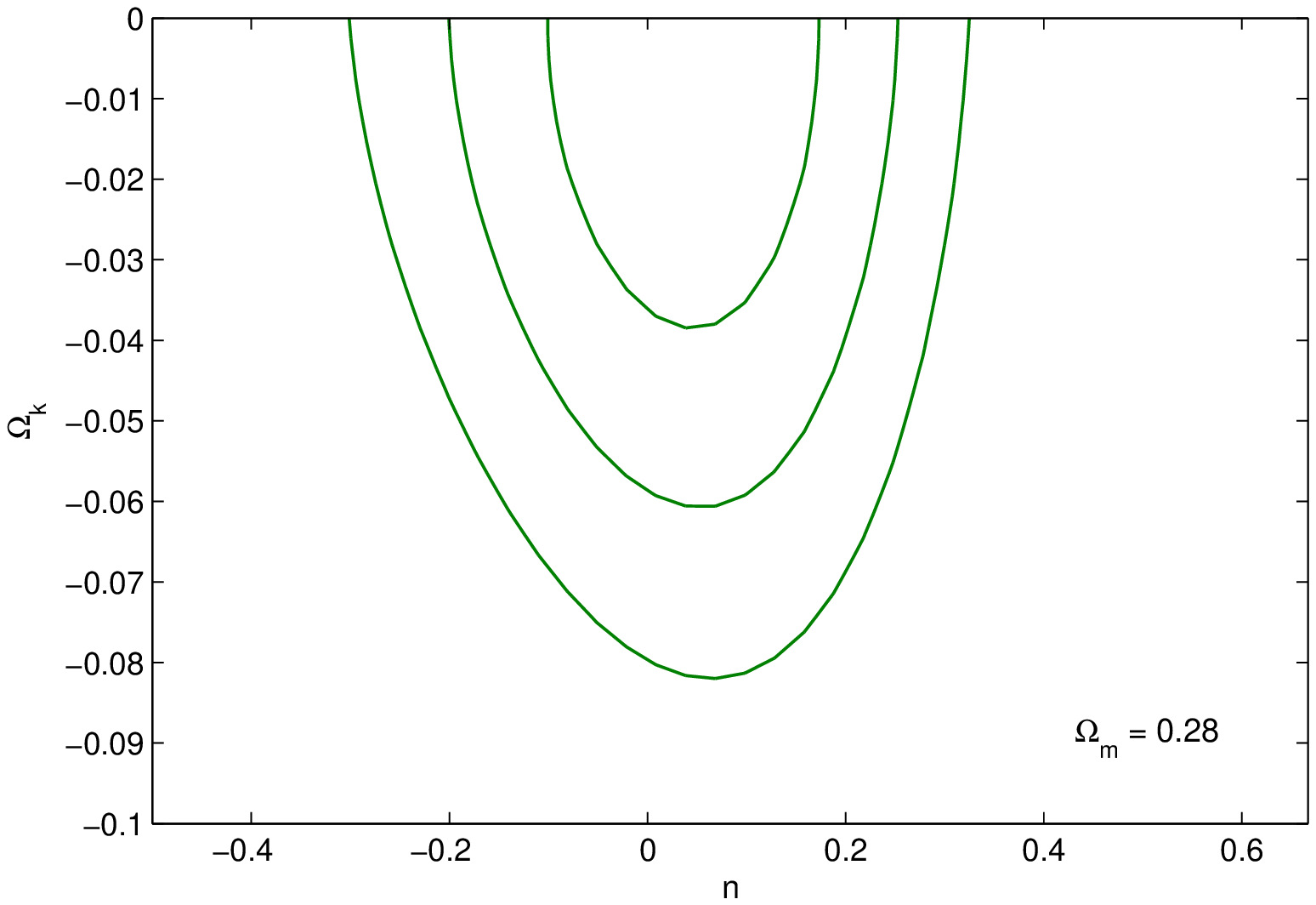}
\includegraphics[height=5.9cm]{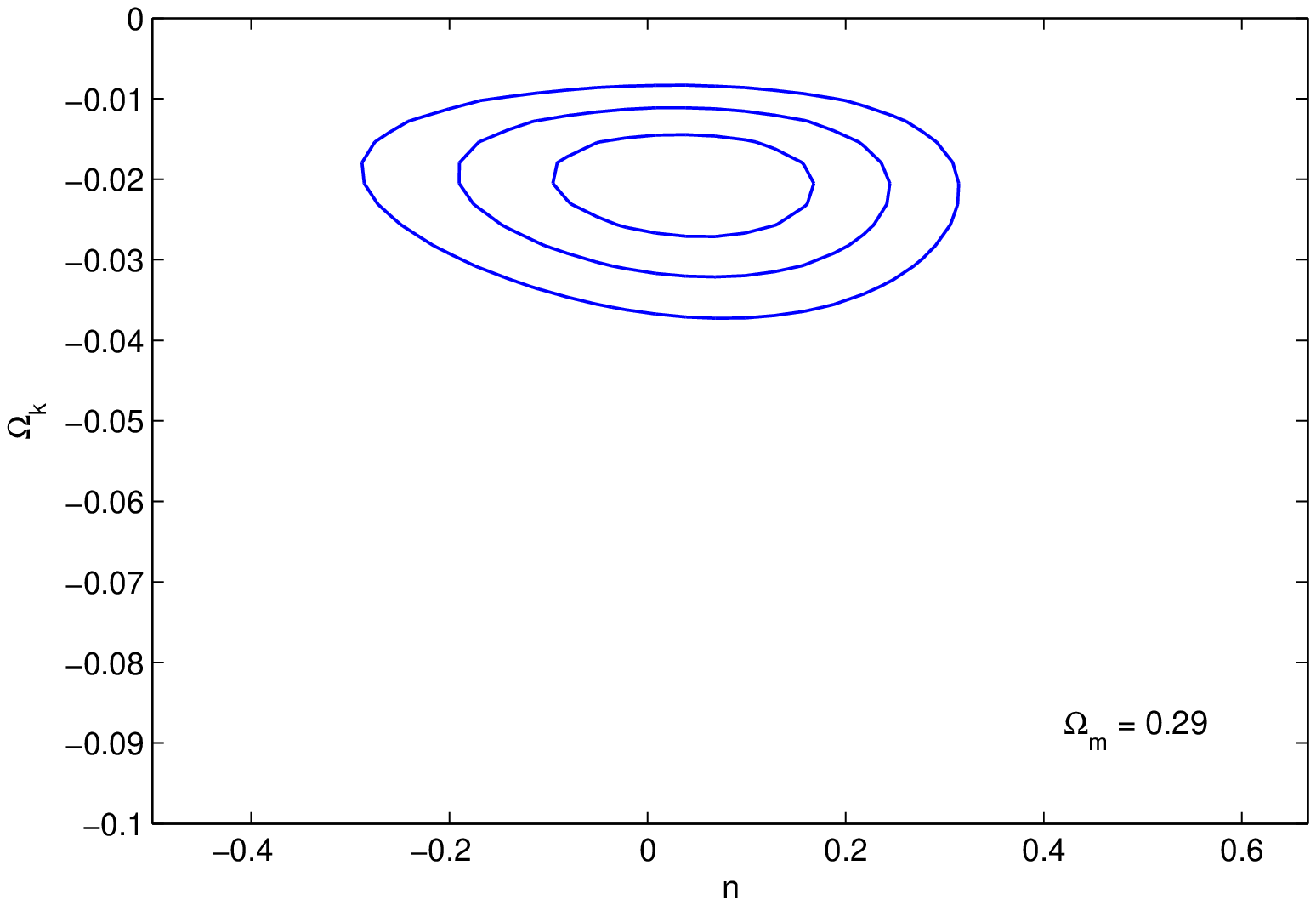}
\caption{\label{Fig:Card}Confidence contours ($68.3\%$, $95.4\%$ and $99.7\%$) for the
Cardassian model, in the $\Omega_m-n$ (top) and $\Omega_k-n$ (bottom) planes,  obtained
with the SN +BAO + CMB data (left) and including also a ${\cal D}$ space topology with
$\gamma=50^\circ \pm 6^\circ$ (right). $\Omega_k$ or $\Omega_m$ are fixed at the value
that minimizes the $\chi^2$.}
\end{center}
\end{figure*}

\subsection{Constraints from cosmic topology}

The effect of cosmic topology is taken into account  by adding a new term to the $\chi^2$
as
\begin{align}
\chi^2_{\rm top}= \left(\dfrac{\chi_{lss}-\chi_{lss}^{\rm th}}{\sigma_{\chi_{lss}}}\right)^2~.
\label{chisqtop}
\end{align}
The value of $\chi_{lss}$ is given by Eq.~(\ref{ChiLSS}) and the
uncertainty $\sigma_{\chi_{lss}}$ comes from the uncertainty $\delta
\gamma$ of the circles-in-the-sky method. The
theoretical value of $\chi_{lss}$ for each model is obtained from
$\chi_{lss}=y(z_{lss})$ combined with the respective expansion law.

\section{ Results}

We have performed a best fit analysis with the minimization of the total $\chi^2$,
\begin{align}
\chi^2=\chi^2_{SN}+\chi^2_{sdss}+\chi^2_{cmb}+\chi^2_{top}~,
\end{align}
for the modified gravity models mentioned above using a MINUIT~\cite{Minuit} based code.
Notice that we marginalize analytically over ${\cal M}'$ and that we  allow the parameters
$\beta$ and $n$ to vary in the interval $]\!-\!10,1\,]\,$ and  $]\!-\!10,2/3\,[\,$,
respectively. In the cases where topology is taken into account, we use the prior
$\Omega_k<0$ since the Poincar\'e dodecahedral space is positively curved. Our results are
summarized in Table~\ref{table:bestfits}.

In Fig. \ref{Fig:LambdaDGP}, we show the 1, 2 and 3 $\sigma$ confidence contours in the
$\Omega_m-\Omega_k$ plane for the $\Lambda$CDM (left) and DGP (right) models obtained using the SNe Ia
gold sample, SDSS acoustic peak  and CMBR shift parameter data (dashed lines). We also show  the
contours obtained assuming a ${\cal D}$ space topology with $\gamma=50^\circ \pm 6^\circ$ (full lines).
This figure together with Table \ref{table:bestfits} shows that the best-fit DGP model is slightly open
whereas the the best-fit $\Lambda$CDM model is slightly closed. Notice also that the effect of the CMBR
shift parameter is to push the best fit values for $\Omega_k$ towards a flat universe, as first pointed
out in Ref. \cite{Maartens:2006yt}, while the SDSS baryon oscillation data constrains basically
$\Omega_m$, in agreement previous works
~\cite{Maartens:2006yt,Guo-Zhu-Alcaniz-Zhang:2006,Fairbairn:2005ue,Alam:2006}.
 Moreover, we see that the effect of including
the topology constraint leads to a reduction of degeneracies, particularly relevant for
the curvature parameter, $\Omega_k$.

Our results for the DT and Cardassian models are shown in Figs.~\ref{Fig:DT}
and~\ref{Fig:Card} both with and without the  topology constraint. The SNe+CMBR+BAO
constraints lead to  best-fit models that are, in both cases, closer to flat than the
best-fit DGP model and closed spaces are preferred. Constraints on $\Omega_m$ are similar
for all models. The effect of topology is, for these models, again clearly important
regarding the curvature parameter, $\Omega_k$, and does not affect significantly the
remaining parameters.
\begin{figure}[htb!]
\begin{center}
\includegraphics[height=6cm]{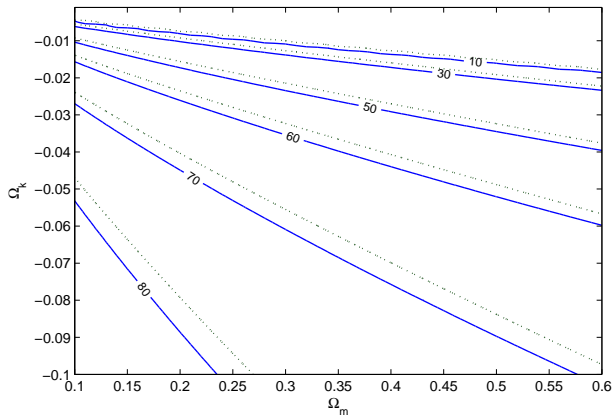}
\caption{Lines of constant angular radius, $\gamma$, in the $\Omega_k-\Omega_m$ plane for
the DGP (solid) and $\Lambda$CDM (dashed) models. } \label{Fig:angular}
\end{center}
\end{figure}

In  Fig.~\ref{Fig:angular} we show how our results regarding the DGP and $\Lambda$CDM
models depend on the angular radius of the circles, $\gamma$. This figure displays the
lines of constant $\gamma$ in the $\Omega_k-\Omega_m$ plane for these models. We see that
when the angular radius increases $\Omega_k$ becomes more negative and that, for any given
value of the angular radius, the $\Lambda$CDM model prefers a less curved universe as
compared with the DGP model. It is also clear that the bounds of topological origin on
$\Omega_k$  are expected to be very similar for both models, for any given angular radius
in the interval $60^\circ \lesssim \gamma \lesssim 10^\circ$; however,  for $\gamma\gtrsim
70^\circ$ the distinction between these bounds for the $\Lambda$CDM and DGP models becomes
more important. Finally, this figure shows that the detection of circles predicted by the
$\mathcal{D}$ topology  restricts substantially the range for $\Omega_k$, regardless of
the angular radius $\gamma$. Clearly additional limits on this density parameter will
arise for a specific value for $\gamma$ and related uncertainty  $\delta \gamma$. We find
that, although the best fit for $\Omega_k$ changes for different values of the angular
radius, $\gamma$, this dependence  is not very strong (see Fig.~\ref{Fig:angular}). For
instance, if we assume a smaller value for $\gamma$, e.g. $\gamma=11^\circ \pm 1^\circ$,
as suggested in Ref.~\cite{Roukema}, the allowed regions will be shifted slightly towards
values of $\Omega_k$ closer to 0. On the other hand, changes in the uncertainty of the
angular radius alter the area corresponding to the confidence regions.

\section{Conclusions}
We have analyzed  observational constraints  on  models that account for the accelerated
expansion of the universe that account for the accelerated expansion of the universe via
long-range corrections to the Friedmann equation, namely the Dvali-Gabadadze-Porrati
braneworld model as well as the Dvali-Turner  and Cardassian models.   Using type Ia
supernovae data together with the baryon acoustic peak in the large scale correlation
function of the Sloan Digital Sky Survey of luminous red galaxies and the Cosmic Microwave
Background Radiation shift parameter data, we find that significant constraints  can be
placed on the parameters of these models.

In general relativity, as well as in any metrical theory of gravitation of some generality
and scope, a common approach to cosmological modeling commences with a space-time manifold
endowed with a Lorentzian metric. The metrical approach to modeling the physical world has
often led physicists to restrict their studies to the purely geometric features of
space-time, either by ignoring the role of spatial topology or by considering just the
simply-connected topological alternatives. However, since the topological properties of a
manifold are more fundamental than its metrical features, it is important to determine to
what extent physical results related to a FLRW universe are constrained by the topology of
its spatial section.

The so-called circles-in-the-sky method makes apparent that a non-trivial detectable
topology of the spatial section of the universe is an observable attribute, and can be
probed for any locally spatially homogeneous and isotropic space.
By assuming the
Poincar\'e dodecahedral space $\mathcal{D}$ as the circles-in-the-sky detected topology of
the spatial sections of the universe, we have re-analyzed the joint SNe+CMBR+BAO
constraints on the abovementioned models. The main outcome of our analysis is that the
detection of a non-trivial spatial topology of the Universe through the circles-in-the-sky
method would give rise to additional constraints on the curvature parameter for the models
we have considered.

\begin{acknowledgments}
M.J.R thanks CNPq for the grant under which this work was carried out, and
is grateful to J.S. Alcaniz for useful discussions. The work of M.C.B. and
O. B. was supported by Funda\c c\~ao para a Ci\^encia e a Tecnologia
(FCT, Portugal) under the grant POCI/FIS/56093/2004.

\end{acknowledgments}

\end{document}